\documentstyle[12pt,epsf]{article}
\newcommand {\beq}{\begin{equation}}
\newcommand {\eeq}{\end{equation}}
\newcommand {\beqa}{\begin{eqnarray}}
\newcommand {\eeqa}{\end{eqnarray}}
\newcommand {\beqan}{\begin{eqnarray*}}
\newcommand {\eeqan}{\end{eqnarray*}}
\newcommand {\n}{\nonumber \\}

\newcommand {\Romannumeral}[1]{\uppercase\expandafter{\romannumeral#1}}

\newcommand {\ee}{\mbox{e}}

\newcommand {\dd}{\mbox{d}}
\newcommand {\sdd}{\mbox{\scriptsize d}}

\newcommand {\del}{\partial}

\begin{document}
\setlength{\oddsidemargin}{0cm}
\setlength{\baselineskip}{7mm}  

\begin{titlepage}
 \renewcommand{\thefootnote}{\fnsymbol{footnote}}
\vspace*{0cm}
    \begin{Large}
       \vspace{2cm}
       \begin{center}
         {\Large Operator Product Expansion \\
in Two-Dimensional Quantum Gravity } \\
       \end{center}
    \end{Large}

  \vspace{1cm}

\begin{center}
           Hajime A{\sc oki}$^{1)}$\footnote
           {
e-mail address : haoki@theory.kek.jp,
{}~JSPS research fellow.},
           Hikaru K{\sc awai}$^{1)}$\footnote
           {
e-mail address : kawaih@theory.kek.jp},
           Jun N{\sc ishimura}$^{2)}$\footnote
           {
e-mail address : nisimura@eken.phys.nagoya-u.ac.jp}{\sc and}
           Asato T{\sc suchiya}$^{1)}$\footnote
           {e-mail address : tsuchiya@theory.kek.jp
}\\
      \vspace{1cm}
        $^{1)}$ {\it National Laboratory for High Energy Physics (KEK),}\\
               {\it Tsukuba, Ibaraki 305, Japan} \\
        $^{2)}$ {\it Department of Physics, 
Nagoya University ,} \\
                 {\it Chikusa-ku, Nagoya 464-01, Japan}\\
\end{center}

\vfill

\begin{abstract}
\noindent 
We consider correlation functions of 
operators and the operator product expansion in two-dimensional quantum gravity.
First we introduce correlation functions with geodesic distances between 
operators kept fixed.
Next by making two of the operators closer, 
we examine if there exists an analog of the operator product expansion 
in ordinary field theories. 
Our results suggest that 
the operator product expansion 
holds in quantum gravity as well,
though special care should be taken regarding the physical meaning of 
fixing geodesic distances on a fluctuating geometry.
\end{abstract}
\vfill
\end{titlepage}
\vfil\eject


\section{Introduction}
\setcounter{equation}{0}
\hspace*{\parindent}
There has been considerable success in the study of 
quantum gravity within the framework of field theory in recent years. 
Particularly, in two dimensions, 
a continuum formalism (Liouville theory) as well as a 
discretized formalism (dynamical triangulation) 
has been consistently developed, 
and they are shown to give equivalent results \cite{KPZ}. 
Correlation functions have been obtained analytically
and they are found to satisfy closed recursive relations, which make the theory
solvable \cite{FKN}. In spite of these developments, 
we are still lacking in the viewpoints of the renormalization group and 
the operator product expansion (OPE), which would provide 
a way to see how the 
theory behaves when we change the scale.
The main difficulty in their realization lies in the fact that in quantum 
gravity the metric field, which could be used to 
fix the scale, is integrated over. 

There have been several attempts to study
the renormalization group in quantum gravity. 
In $2+\epsilon$-dimensional quantum gravity, the renormalization point can 
be introduced as in the ordinary perturbation theory, 
and the dependence of the coupling constants on the renormalization point has 
been studied \cite{epsilon}.
Block-spin transformation has been considered in the context of 
dynamical triangulation by 
various people, and numerical studies seem to support the validity of the 
formalism at least in two dimensions \cite{block}. 
Also there is a study on the effects of the 
gravitational dressing to the renormalization group
in two-dimensional quantum gravity \cite{KKP}.

However, the OPE in quantum gravity has not been studied yet.
In this paper, we study it in two-dimensional quantum gravity. 
For this purpose, we must introduce the notion of 
the distance between local operators. 
In quantum gravity we can use the geodesic distance, 
which is general coordinate 
invariant. 
Recently a formalism has been developed, which enables us 
to introduce the geodesic distance \cite{KKMW,IK}. 
Using this formalism we calculate correlation functions with fixed geodesic 
distances between the operators. 
By making two of the operators closer to each other, we examine if there exist
OPE like relations among the operators. 
We insert other operators as observers and compare the correlation functions.
However, since the metric is fluctuating in quantum gravity, 
it is not trivial if fixing the geodesic 
distance corresponds exactly to fixing the scale as in the ordinary quantum
field theory in the flat space. 
The OPE is expected to hold as a result of integrating out the local degrees 
of freedom. 
In quantum gravity, however, the meaning of the local degrees of freedom 
is somewhat obscure, since the metric itself is the dynamical variable. 
The observers, each of which we require to be at a definite distance 
from the two close operators, 
might be sensitive 
to some large fluctuations of ``the local degrees 
of freedom'' which should have been integrated out. 
We would like to shed light on such a subtlety in 
fixing geodesic distance in quantum gravity.


Another interesting aspect we can elucidate by using this kind of
correlation functions 
is the fractal structure 
of the space-time, which 
has been revealed in two dimensions in Ref. \cite{KKMW}.
There, sections of the two-dimensional surface were considered, 
each of which was 
at a fixed geodesic distance 
from a given point. 
A typical section is composed of loops of various lengths, whose distribution 
could be calculated analytically. 
The loop-length distribution shows a scaling behavior, which can be interpreted 
as the fractal structure of space-time. 
However, it was found that the total length of the section at a fixed geodesic 
distance is divergent, and in this sense the fractal structure 
can only be seen in a somewhat indirect way through the loop-length 
distribution. Here 
we show that the two-point functions of the cosmological constant terms with 
fixed geodesic distances provides a more direct way to see 
the fractal structure.

This paper is organized as follows. 
In Section 2, we introduce two-point functions with fixed geodesic 
distances.
In Section 3, we show how we can see the fractal structure of the space-time 
in a direct way using these two-point functions.
In Section 4, we calculate three-point functions with fixed geodesic distances.
By using them, 
together with the results for one-point and two-point functions, 
we examine whether there are OPE like operator relations or not.
Section 5 is devoted to the summary and outlook. 

\vspace{1cm}

\section{Two-point Functions with Fixed Geodesic Distances}
\begin{figure}
\begin{center}
\leavevmode
\epsfbox{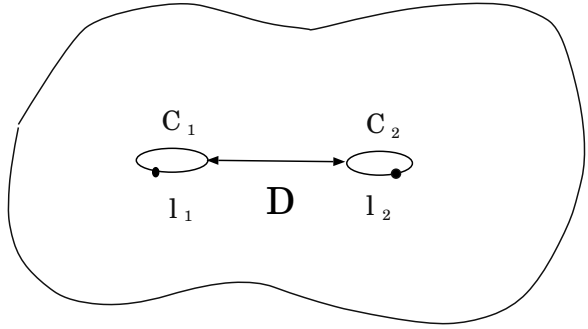}
\caption{The correlation function of two loops  
with fixed geodesic distance $D$. 
The geodesic distance between the loop $C_{1}$ of length $l_{1}$ 
and the loop $C_{2}$ of length $l_{2}$ is fixed to be $D$.
Each of the loops is considered to have a marking point. }
\label{fig:imo}
\end{center}
\vspace{1cm}
\end{figure}

\setcounter{equation}{0}
\hspace*{\parindent}
Throughout this paper, we consider pure gravity in 
two dimensions.\footnote[1]{Since we treat pure gravity we cannot compare 
results of our 
calculations with those of any theories in flat space.
It is interesting to include matter degrees of freedom
and see how OPE in ordinary theories in flat space is modified 
due to the effects of coupling to gravity.}
For simplicity, the topology of the two-dimensional manifold 
is restricted to be a sphere.
We introduce a correlation function of two loops with fixed geodesic distance, 
which is formally defined as follows (See Fig. \ref{fig:imo}).
\beqa
G(l_1,l_2;D;t)
&=&
\int \frac{{\cal D}g}{\mbox{Vol(Diff)}}
~\delta( \int_{C_1} \sqrt{g_{\mu\nu}\dd x^{\mu}\dd x^{\nu}} - l_1)
~\delta( \int_{C_2} \sqrt{g_{\mu\nu}\dd x^{\mu}\dd x^{\nu}} - l_2) \n
&~&\delta(d(C_1,C_2)-D) 
~\ee^{-t \int \sdd^2x \sqrt{g}}.
\label{eq:twoloop}
\eeqa
The geodesic distance $d(C_1,C_2)$ between the loops 
$C_1$ and $C_2$ is defined by 
\beq
d(C_1,C_2)= \min_{P_1 \in C_1,P_2 \in C_2} d(P_1,P_2),
\eeq
where $d(P_1,P_2)$ is the geodesic distance between the points 
$P_1$ and $P_2$ in the ordinary sense. 
Here we consider such an amplitude that 
each of the loops $C_1$ and $C_2$ has a marking point.

This quantity can be calculated using the proper-time evolution kernel,
which is defined as the amplitude with the entrance loop $C$ of length $l$ 
and the exit loop of length $l'$, 
where each point $P'$ on the exit loop $C'$ is at the geodesic distance $D$ 
from the entrance loop $C$ in the sense that
\beq
\min_{P\in C} d(P',P)= D.
\eeq
We adopt the convention that  
the exit loop has a marking point and the entrance loop not.
The Laplace transform of the proper-time evolution kernel is 
obtained as \cite{KKMW}
\beq
N(\zeta,\zeta';D;t) =
\frac{1}{\zeta'+ A(\zeta;D;t)}.
\eeq
$A(\zeta;D;t)$ is a function defined as 
\beq
A(\zeta;D;t) = f(x f^{-1}(\zeta))  ,
\eeq
where 
\beqa
x&=& \ee^{\sqrt{6} ~ t^{1/4} D }, \\
f(y) &=&  \sqrt{t} ~ \left\{ 
\frac{3}{2} \left(
\frac{2}{y-1}+1
\right)^2 -1
 \right\},  \\
f^{-1}(\zeta) &=& 
\frac{\sqrt{6}~ t^{1/4}}
{\sqrt{\zeta+ \sqrt{t}}-\sqrt{\frac{3}{2}} ~ t^{1/4} } +1.
\eeqa
\begin{figure}
\begin{center}
\leavevmode
\epsfbox{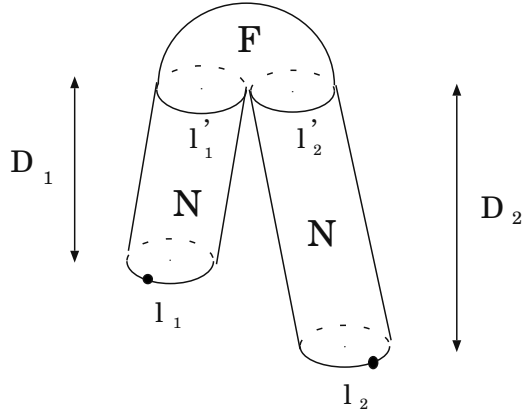}
\caption{Calculation of the correlation function of two loops
$G(l_{1},l_{2};D;t)$.
         The diagram corresponds to Eq. (\protect \ref{eq:twopoint}).
         The correlation function of two loops can thus be calculated 
using the disk amplitude $F$
         and the proper-time evolution kernel $N$.}
\label{fig:mata}
\end{center}
\vspace{1cm}
\end{figure}

Using the proper-time evolution kernel, we can express the 
correlation function of two loops $G(l_1,l_2;D;t)$ as
\beq
G(l_1,l_2;D;t)
=
l_1 ~l_2 ~\int_0^{\infty} \dd l_1 ' 
\int_0^{\infty} \dd l_2 ' 
~F(l_1' + l_2';t) 
~N(l_1,l_1';D_1;t)
~N(l_2,l_2';D_2;t), 
\label{eq:twopoint}
\eeq
where $F(l;t)$ is the disc amplitude, whose Laplace transform 
is given by 
\beq
F(\zeta;t)= (2 \zeta - \sqrt{t}) \sqrt{\zeta + \sqrt{t}},
\eeq
and $D = D_1 + D_2$.
Fig. \ref{fig:mata} indicates the diagram corresponding to 
Eq. (\ref{eq:twopoint}).
Eq. (\ref{eq:twopoint}) 
can be evaluated as 
\beqa
G(\zeta_1,\zeta_2;D;t)
&=&
\frac{\del}{\del \zeta_1}~
\frac{\del}{\del \zeta_2} ~
\int \frac{\dd \zeta }{2 \pi i}
~F(\zeta;t) 
~N(\zeta_1,-\zeta;D_1;t)
~N(\zeta_2,-\zeta;D_2;t) \\
&=&
- \frac{\del}{\del \zeta_1}~
\frac{\del}{\del \zeta_2} ~
\frac{F(A_1)-F(A_2)}{A_1-A_2}  \\
&=&
- 2 \sqrt{6} ~ t^{1/4}
\frac{1+y_1 y_2}{(1-y_1 y_2)^3}
\frac{\del y_1}{\del \zeta_1}
\frac{\del y_2}{\del \zeta_2}, 
\label{eq:twoloopresult}
\eeqa
where
\beqa
A_i &=& A(\zeta_i;D_i;t), \\
y_i &=& x_i f^{-1}(\zeta_i), \\
x_i &=& \ee ^{\sqrt{6}  ~ t^{1/4} D_i } .
\eeqa
The r.h.s of (\ref{eq:twoloopresult})
should depend only on $D=D_1 + D_2$, and this is indeed the case 
since it is written in terms of 
$y_1 y_2$, which is 
\beqa
y_1 y_2 &=& x_1 x_2 f^{-1}(\zeta_1) f^{-1}(\zeta_2) \\
        &=& \ee ^{\sqrt{6} ~ t^{1/4} (D_1 +D_2)  } 
f^{-1}(\zeta_1) f^{-1}(\zeta_2). 
\eeqa
This is an example of the consistency condition mentioned in 
Ref. \cite{IK2}.

Let us next pinch the loops in order to obtain the local operators. 
This can be done by expanding the above result in terms of 
$1/\sqrt{\zeta_1}$ and $1/\sqrt{\zeta_2}$ as 
\beq
G(\zeta_1,\zeta_2;D;t) =
\sum_{n,m=1}^{\infty} G_{nm}(D,t) 
~\zeta_{1}^{-(\frac{n}{2}+1)}
\zeta_{2}^{-(\frac{m}{2}+1)},
\eeq
where $G_{nm}(D,t) $ can be regarded as the two-point correlation 
function of the operators ${\cal O}_n$ and ${\cal O}_m$ with 
the fixed geodesic distance $D$.
\beq
G_{nm}(D;t)= \langle {\cal O}_n {\cal O}_m (D) \rangle. 
\eeq
The explicit forms of $G_{nm}$'s are obtained as
\beqan
G_{11}(D,t) &=& 3 \sqrt{6} ~ t^{3/4} ~ \frac{(x+1)~x}{(x-1)^3}, \\
G_{12}(D,t) &=&  - 18 ~ t~ \frac{(x^2+4x+1)~x}{(x-1)^4}, \\
G_{22}(D,t) &=&  18 \sqrt{6}~ t^{5/4}~ \frac{(x+1)(x^2+10x +1)~x}{(x-1)^5},\\
&\vdots&
\eeqan
where $x=\ee ^{\sqrt{6} ~ t ^{1/4} D  }$.
Note that when one considers correlation functions of loops without fixing the 
geodesic distances and expands them in terms of $1/\sqrt{\zeta_i}$, 
one encounters only odd powers in contrast to the above results 
where we encounter even powers as well.
If we integrate $G_{nm}(D,t) $ over $D$ from 0 to the infinity, 
we reproduce the conventional results of the two-point functions. 
Indeed the cases including at least one even number index 
vanish after the integration. 
This means that we have found a new set of operators ${\cal O}_n$ with even $n$,
which cannot be seen without fixing the geodesic distance. 
We also comment that $G_{11}(D,t)$ agrees with the result obtained in 
Ref. \cite{AW} in a different way.

\vspace{1cm}

\section{Fractal Structure of the Space-time} \label{sec:fractal}
\setcounter{equation}{0}
\hspace*{\parindent}
Using the two-point correlation functions with fixed geodesic 
distances obtained in the previous 
section, we can study the fractal structure of the space-time. 
For this purpose, let us consider the two-point correlation function 
of the cosmological constant terms, which can be formally written in the 
following way.
\beq
V'(D)=
\lim_{A \rightarrow \infty}
\frac{1}{A}
\frac
   {  \int \frac{{\cal D}g}{\mbox{\scriptsize Vol(Diff)}} 
      ~ \int \dd^2 x_1 \sqrt{g(x_1)} 
      ~\int \dd^2 x_2 \sqrt{g(x_2)}
      ~\delta(d(x_1,x_2)-D)
      ~\delta(\int \sqrt{g} \dd ^2 x - A )
   }
   {  \int \frac{{\cal D}g}{\mbox{\scriptsize Vol(Diff)}} 
~ \delta(\int \sqrt{g} \dd ^2 x - A )
   }.
\label{eq:hausdorff}
\eeq
This quantity can be identified with 
the $D$-derivative of the volume of the region within 
the geodesic distance $D$ from a given point.
Therefore, if $V'(D)$ behaves as $D^{d_h-1}$,
we can identify $d_h$ as the Hausdorff dimension of the space-time. 

The thermodynamic limit $A \rightarrow \infty$ can be obtained by 
expanding Laplace-transformed expressions around $t=0$.  
If we expand $F(\zeta;t)$ in terms of $t$,
we have
\beqa
F(\zeta;t) &=&
 \sum_{k \geq -5;\mbox{\scriptsize odd}}
 F_k(t) ~ \zeta ^{-(\frac{k}{2}+1)}  \n
&=&
2 ~ \zeta^{3/2} - \frac{3}{4} ~\zeta^{1/2}~ t 
+ \frac{1}{4}~ \zeta ^{-3/2}~ t ^{3/2} + O(t^2)  .
\label{eq:FkTD}
\eeqa
Since the first two terms come from small universes, they should be subtracted 
when we take the thermodynamic limit $t \rightarrow 0$.
Therefore the only one-point function 
that remains nonzero after the thermodynamic limit
is $F_1$.
As in the case of two-point functions, we identify  $F_1$ as an expectation
value of the scaling operator ${\cal O}_1$, which is known to correspond to
the cosmological constant term.
In what follows, we use the notation
\beqa
F_1 ^{\mbox{\scriptsize (T.D.)}} &=& \frac{1}{4}~t^{3/2},  \\
F_k ^{\mbox{\scriptsize (T.D.)}} &=& 0 ~~~~~~~~~(k \neq 1) .
\label{eq:F1}
\eeqa

Similarly if we expand $G_{11}$  in terms of $t$,
we have
\beq
G_{11}(D;t) = \frac{1}{D^3} - \frac{3}{20}~ D~ t 
+ \frac{1}{14}~ D^3~ t^{3/2} + O(t^2).
\eeq
Hence in the thermodynamic limit, we have 
\beq
G_{11}^{\mbox{\scriptsize (T.D.)}} (D)=
 \frac{1}{14}~ D^3~t^{3/2}. 
\eeq
Here, we have two terms $O(t^0)$ and $O(t^1)$ that have to be subtracted 
when we take the $t\rightarrow 0$ limit, 
which is in contrast to the case of 
the usual cylinder amplitude, where we need only one subtraction. 
This can be naturally understood, 
since the two operators, which are constrained to be apart from 
each other by the distance $D$, behave as one operator in a large space-time. 

Using the above expressions in the thermodynamic limit, 
(\ref{eq:hausdorff}) can be evaluated as
\beq
V'(D)=
\frac
   {
      G_{11}^{\mbox{\scriptsize (T.D.)}}
   }
   {
      F_{1}^{\mbox{\scriptsize (T.D.)}}
   }
=\frac{2}{7}~D^3.
\eeq

Therefore, 
the Hausdorff dimension of the space-time is 
$d_h=4$, which agrees with the one given through 
the loop-length distribution \cite{KKMW}.
In Ref. \cite{KKMW}, it has been pointed out that 
the total length of the boundary at the geodesic distance $D$ 
from a given point is not well-defined in the continuum limit. 
However, the area of the region within the geodesic distance 
$D \sim D + \Delta D$ is a well-defined quantity as we have seen.
Thus, we are able to observe
the fractal structure of the space-time more directly 
by considering the two-point correlation functions of the 
cosmological constant terms than by considering the loop-length distribution. 

We note that in Ref. \cite{AW} it is also argued that $d_{h}=4$
by examining the scaling behavior for the finite volume universe.

\vspace{1cm}

\section{Operator Product Expansion in Quantum Gravity}
\setcounter{equation}{0}
\hspace*{\parindent}
In this section, we examine if the operator product expansion holds
in quantum gravity 
in the sense that two operators close to each other can be 
regarded as one when viewed from a distance. 
Although we are able to obtain some of the OPE coefficients by 
comparing two-point functions and one-point functions, 
the first nontrivial check of the OPE as operator relations 
can be given by inserting one extra operator as an observer. 
For this purpose, we need the three-point functions,
which we calculate in the next subsection.

\subsection{Three-Point Functions with Fixed Geodesic Distances}

\begin{figure}
\begin{center}
\leavevmode
\epsfbox{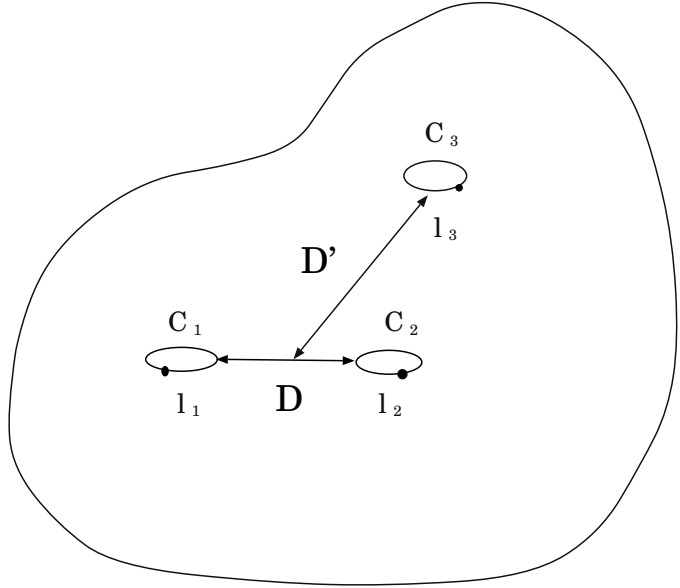}
\caption{The correlation function of three loops with fixed geodesic distances.
         The geodesic distance between the loop $C_{1}$ 
of length $l_{1}$ and 
         the loop $C_{2}$ of length $l_{2}$ is fixed to be $D$.
         The geodesic distance between the loop $C_{3}$ 
of length $l_{3}$ and 
     the union of the loops $C_{1}$ and $C_{2}$ is fixed to be $D'$.
         Each of the loops is considered to have a marking point.} 
\label{fig:okarina}
\end{center}
\vspace{1cm}
\end{figure}

We consider a correlation function of three loops 
with fixed geodesic distances, which is  
formally defined as follows (See Fig. \ref{fig:okarina}).
\beqa
H(l_1,l_2,l_3;D;D';t)
&=&
\int \frac{{\cal D}g}{\mbox{Vol(Diff)}}
\prod_{i=1}^3 ~ \delta( \int_{C_i} \sqrt{g_{\mu\nu}
\dd x^{\mu}\dd x^{\nu}} - l_i) \n
&~& \delta(d(C_1,C_2)-D) ~\delta(d(C_1 \cup C_2 ,C_3)-D') 
~\ee^{-t \int \sdd^2 x \sqrt{g}},
\label{eq:threeloop}
\eeqa
where we have specified the geodesic distance between $C_3$ and 
the union of $C_1$ and $C_2$. 
Each of the three loops is considered to have a marking point.

\begin{figure}
\begin{center}
\leavevmode
\epsfbox{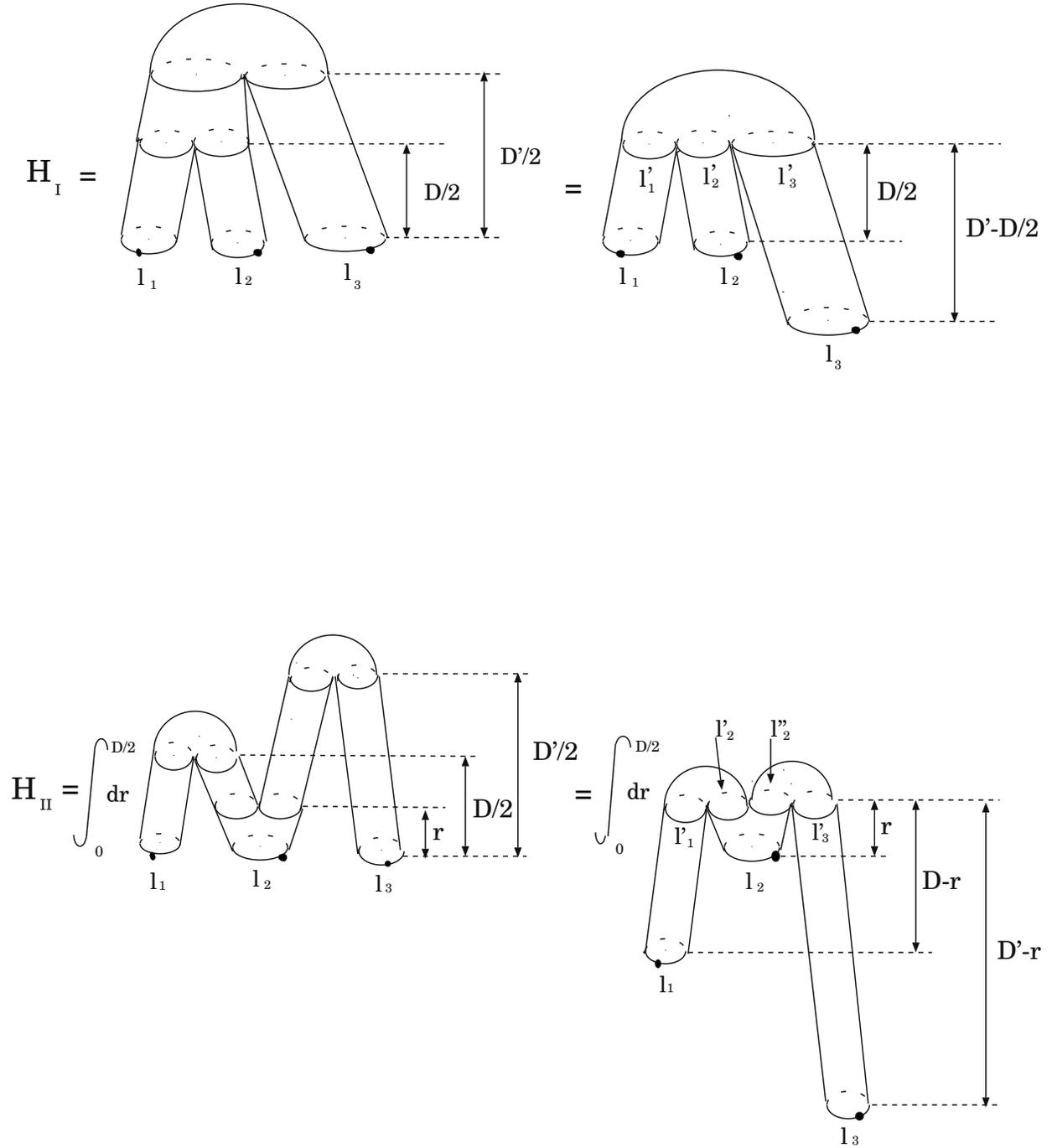}
\caption{Calculation of the correlation function of three loops 
$H(l_{1},l_{2},l_{3};D;D';t)$.
         The diagrams in the r.h.s. correspond to Eq. (\protect \ref{eq:H1}) 
         and Eq. (\protect \ref{eq:H2}) respectively.
         The equalities of the diagrams come from the consistency condition.}
\label{fig:mitsumata}
\end{center}
\vspace{1cm}
\end{figure}

When $D < D'$, we can calculate the above quantity by the same 
technique with which we calculate the correlation functions of two loops. 
$H(l_1,l_2,l_3;D;D';t)$ can be calculated as the sum of three 
contributions. 
\beq
H(l_1,l_2,l_3;D;D';t)=
H_{\mbox{\scriptsize \Romannumeral{1}}} (l_1,l_2,l_3;D;D';t)
+H_{\mbox{\scriptsize \Romannumeral{2}}} (l_1,l_2,l_3;D;D';t)
+H_{\mbox{\scriptsize \Romannumeral{2}}}' (l_1,l_2,l_3;D;D';t).
\eeq
$H_{\mbox{\scriptsize \Romannumeral{1}}} $ and 
$H_{\mbox{\scriptsize \Romannumeral{2}}} $ 
correspond to the two diagrams in the l.h.s. of Fig. \ref{fig:mitsumata} 
respectively. 
$H_{\mbox{\scriptsize \Romannumeral{2}}}' $ 
can be obtained by exchanging $l_1$ and $l_2$ in 
$H_{\mbox{\scriptsize \Romannumeral{2}}} $.

Making use of the consistency condition, we can evaluate 
$H_{\mbox{\scriptsize \Romannumeral{1}}} $ and 
$H_{\mbox{\scriptsize \Romannumeral{2}}} $ by the 
diagrams in the r.h.s. of Fig. \ref{fig:mitsumata}, 
which can be expressed as
\beqa
H_{\mbox{\scriptsize \Romannumeral{1}}} (l_1,l_2,l_3;D;D';t)
&=& 
l_1  ~ l_2 ~ l_3~
\int _0^{\infty} \dd l_1 ' ~ \dd l_2 '~ \dd l_3 '
  ~ (l_1' + l_2' ) ~  F(l_1 ' + l_2 ' + l_3 ') \n
&~&N(l_1,l_1 ';\frac{D}{2}; t)
~N(l_2,l_2 ';\frac{D '}{2}; t)
~N(l_3,l_3 ';D'- \frac{D}{2}; t),
\label{eq:H1}
\eeqa
\beqa
H_{\mbox{\scriptsize \Romannumeral{2}}} (l_1,l_2,l_3;D;D';t)
&=& 
l_1 ~ l_2  ~ l_3 ~
\int _0^{D/2} \dd r
\int _0^{\infty} \dd l_1 ' ~ \dd l_2 ' ~ \dd l_2 '' ~ \dd l_3 '
 ~  l_2' ~  l_2''  
~ F(l_1 ' + l_2 ') ~ F(l_2 '' + l_3 ')  \n
&~&N(l_1,l_1 ';D - r ; t)
~N(l_2,l_2 '+ l_2 ''; r ; t)
~N(l_3,l_3 '; D'- r; t).
\label{eq:H2}
\eeqa
We obtain the following results for 
$H_{\mbox{\scriptsize \Romannumeral{1}}} $ and 
$H_{\mbox{\scriptsize \Romannumeral{2}}} $.
\beqa
&~& H_{\mbox{\scriptsize \Romannumeral{1}}} (\zeta_1,\zeta_2,\zeta_3;D;D';t) \n
&=& 
\frac{\del}{\del \zeta_1}  ~ \frac{\del}{\del \zeta_2} ~ 
\frac{\del}{\del \zeta_3}~
\left[
      F'(A_1)\frac{1}{A_1-A_2}\frac{1}{A_1-A_3}
    + F'(A_2)\frac{1}{A_2-A_1}\frac{1}{A_2-A_3}    \right.  \n
&~&  - F(A_1)\frac{1}{A_1-A_2}\frac{1}{(A_1-A_3)^2} 
    - F(A_2)\frac{1}{A_2-A_1}\frac{1}{(A_2-A_3)^2}  \n
&~& \left. + F(A_3)\frac{1}{(A_3-A_1)^2}\frac{1}{A_3-A_2}
    + F(A_3)\frac{1}{(A_3-A_2)^2}\frac{1}{A_3-A_1}
\right],
\eeqa
where $A_i=A(\zeta_i;D_i;t)$ with
\beqa
D_1 &=& \frac{D}{2}, \\
D_2 &=& \frac{D}{2}, \\
D_3 &=& D'-\frac{D}{2}.
\eeqa
\beqa
&~& H_{\mbox{\scriptsize \Romannumeral{2}}} (\zeta_1,\zeta_2,\zeta_3;D;D';t) \n
&=& 
- \frac{\del}{\del \zeta_1}  ~ \frac{\del}{\del \zeta_2} ~ 
\frac{\del}{\del \zeta_3}~
\int_0^{D/2} \dd r
\left[
      \frac{1}{2} F'(A_2)^2 \frac{1}{A_1-A_2}\frac{1}{A_3-A_2} \right.  \n
&~&    + F(A_2) F'(A_2)\frac{1}{(A_1-A_2)^2}\frac{1}{A_3-A_2}   
  - F(A_1) F'(A_2)\frac{1}{(A_1-A_2)^2}\frac{1}{A_3-A_2}   \n
&~&    + \frac{1}{2} F(A_2)^2 \frac{1}{(A_1-A_2)^2}\frac{1}{(A_3-A_2)^2}  
 - F(A_1)F(A_2)\frac{1}{(A_1-A_2)^2}\frac{1}{(A_3-A_2)^2}  \n
&~& \left.
    + \frac{1}{2} F(A_1)F(A_3)\frac{1}{(A_1-A_2)^2}\frac{1}{(A_3-A_2)^2} 
+ (1 \leftrightarrow 3)
\right],
\eeqa
where $A_i=A(\zeta_i;D_i;t)$ with 
\beqa
D_1 &=& D-r, \\
D_2 &=& r, \\
D_3 &=& D'-r.
\eeqa

In order to calculate the three-point functions, 
we expand $H(\zeta_1,\zeta_2,\zeta_3;D;D';t)$ in terms of 
$1/\sqrt{\zeta_1}$, $1/\sqrt{\zeta_2}$ and $1/\sqrt{\zeta_3}$ as
\beq
H(\zeta_1,\zeta_2,\zeta_3;D;D';t)
=
\sum _{n,m,l = 1}^{\infty} H_{nml} (D;D';t)
~\zeta_1 ^{- (\frac{n}{2}+1)}
\zeta_2 ^{- (\frac{m}{2}+1)}
\zeta_3 ^{- (\frac{l}{2}+1)}.
\eeq
We can identify $H_{nml} (D;D';t)$ as the three-point function
of the operators ${\cal O}_n$, ${\cal O}_m$ and ${\cal O}_l$ with the 
fixed geodesic distance $D$ between the first two and with the fixed 
geodesic distance $D'$ between the union of the first two and 
the third one.

\subsection{Calculation of the Coefficients of the OPE}

Let us study if there exists a set of operator relations, which corresponds
to the OPE, such as
\beq
{\cal O}_n  {\cal O}_m (D) \sim 
\sum _{k=1} ^{\infty} 
C_{nm}^{~~~k} D^{4-n-m+k} {\cal O}_k 
\label{eq:ope}
\eeq
when $D \rightarrow 0$. 
The power of $D$ in the r.h.s. can be determined through dimensional 
analysis by considering that 
$[{\cal O}_n]= M^{\frac{n-5}{2}}$ \cite{FKN} and $[D]= M^{-\frac{1}{2}}$ 
\cite{KKMW,IK} and that fixing the geodesic distance $D$ 
 by delta function in the l.h.s. gives $[D^{-1}] \sim M^{\frac{1}{2}}$.

If there is such an identity between operators, 
we should have the following relations.
\beq
G_{nm}^{\mbox{\scriptsize (T.D.)}}(D)
=
\sum _{k=1} ^{\infty} 
C_{nm}^{~~~k} D^{4-n-m+k} F_k ^{\mbox{\scriptsize (T.D.)}},  
\label{eq:twoandone}
\eeq
\beq
H_{nml}^{\mbox{\scriptsize (T.D.)}}(D,D')
=
\sum _{k=1} ^{\infty} 
C_{nm}^{~~~k} D^{4-n-m+k} G_{kl} ^{\mbox{\scriptsize (T.D.)}}(D').  
\label{eq:threeandtwo}
\eeq
Note that from dimensional analysis it follows that 
\beq
G_{nm} ^{\mbox{\scriptsize (T.D.)}}(D) \propto D^{5-n-m}.
\label{eq:GnmTD}
\eeq

Because of (\ref{eq:GnmTD}) and (\ref{eq:F1}), we can determine 
$C_{nm}^{~~~k}$ only for $k=1$  through (\ref{eq:twoandone}) 
by comparing the two-point functions and the one-point function. 
The results are shown in the second column of 
Tables \ref{tab:C11k}-\ref{tab:C22k}.
\begin{table}
\begin{center}
\caption{The results for $C_{11}^{~~k}$.
            We show the result from the comparison between the 
two-point function
            $G_{11}$ and the one-point function $F_{1}$ in the second column, 
            and those from the comparison between the three-point functions 
            $H_{11l}$ and the two-point functions $G_{kl}$ in the third column.
            Note that in the latter case, we can use various kinds of 
            observer operators $O_{l}$ labeled by $l$.}
  \label{tab:C11k}
 \vspace{5mm}
\def\arraystretch{1.7}
  \begin{tabular}{|c|c|c|c|c|c|c|}  \hline
    &2-point and&\multicolumn{5}{c|}{3-point and 2-point}\\ \cline{3-7}
k       & 1-point &$ l=1\sim 6$ & $l=7$ & $l=8$ & $l=9$ & $l=10$ \\ \hline
1         &$\frac{2}{7}$&$ \frac{2}{7} $&$ \frac{2}{7} $&$ \frac{2}{7} $&$ \frac{2}{7}$&$ \frac{2}{7}$ \\ \hline
2         &-    &$\frac{ 205}{1792 }$&$\frac{ 205}{1792 }$&$\frac{ 205}{1792 }$&$\frac{ 205}{1792 }$&$\frac{ 205}{1792}$ \\ \hline
3         &-       &$\frac{ 3}{128 }$&$\frac{ 3}{128 }$&$\frac{ 3}{128 }$&$\frac{ 3}{128 }$&$\frac{ 3}{128}$ \\ \hline
4         &-       &$\frac{ 1}{152 }$&$\frac{ 1}{152 }$&$\frac{ 1}{152 }$&$\frac{ 1}{152 }$&$\frac{ 1}{152}$ \\ \hline
5         &-       &- &0 &0 &0 &0 \\ \hline
6        &-       &- &0 &0 &0 &0 \\ \hline
7         &-   &$\frac{ 20}{637 }$&$\frac{ 10}{637 }$&$\frac{ 20}{5733 }$&$\frac{ 20}{23569 }$&$\frac{ 20}{77077}$ \\ \hline
8         &- &$\frac{ 9647}{1490944 }$&$\frac{ 9647}{1677312 }$&$\frac{ 9647}{2981888 }$&$\frac{ 877}{745472 }$&$\frac{ 9647}{23855104}$ \\ \hline
9         &- &$\frac{ 1}{1536 }$&$\frac{ 3}{4736 }$&$\frac{ 3}{5632 }$&$\frac{ 1}{3072 }$&$\frac{ 1}{6656}$ \\ \hline
10        &- & $\frac{ 1}{40960 }$&$\frac{ 3}{123904 }$&$\frac{ 3}{131072 }$&$\frac{ 1}{53248 }$&$\frac{ 1}{81920}$ \\ \hline
11        &- &0 &0 &0 &0 &0 \\ \hline
12        &- &0 &0 &0 &0 &0 \\ \hline
13        &- &0 &0 &0 &0 &0 \\ \hline
14        &- &0 &0 &0 &0 &0 \\ \hline
15        &- &0 &0 &0 &0 &0 \\ \hline 
    \end{tabular}
\end{center} 
\end{table}

\begin{table}
\begin{center}
   \caption{The results for $C_{12}^{~~k}$.
            We show the result from the comparison between the 
two-point function
            $G_{12}$ and the one-point function $F_{1}$ in the second column, 
            and those from the comparison between the three-point functions 
            $H_{12l}$ and the two-point functions $G_{kl}$ in the third column.
            Note that in the latter case, we can use various kinds of 
            observer operators $O_{l}$ labeled by $l$.}
  \label{tab:C12k}
\vspace{5mm}
\def\arraystretch{1.7}
   \begin{tabular}{|c|c|c|c|c|c|c|}  \hline
    &2-point and&\multicolumn{5}{c|}{3-point and 2-point}\\ \cline{3-7}
$k$ & 1-point & $l=1\sim 6$&$l= 7 $&$l= 8 $&$l= 9 $&$l= 10$ \\ \hline
1         &$\frac{6}{7      }$&$\frac{6}{7  }$&$\frac{6}{7  }$&$\frac{6}{7  }$&$\frac{6}{7  }$&$\frac{6}{7}$ \\ \hline
2         &-        &$\frac{269}{448 }$&$\frac{269}{448 }$&$\frac{269}{448 }$&$\frac{269}{448 }$&$\frac{269}{448}$ \\ \hline
3         &-        &$\frac{415}{1792 }$&$\frac{415}{1792 }$&$\frac{415}{1792 }$&$\frac{415}{1792 }$&$\frac{415}{1792}$ \\ \hline
4         &-        &$\frac{3}{64 }$&$\frac{3}{64 }$&$\frac{3}{64 }$&$\frac{3}{64 }$&$\frac{3}{64}$ \\ \hline
5         &-        &- &$\frac{1}{256 }$&$\frac{1}{256 }$&$\frac{1}{256 }$&$\frac{1}{256}$ \\ \hline
6         &-        &- &0 &0 &0 &0 \\ \hline
7         &-        &$\frac{180}{637 }$&$\frac{90}{637 }$&$\frac{20}{637 }$&$\frac{180}{23569 }$&$\frac{180}{77077}$ \\ \hline
8         &-       &$\frac{58475}{745472 }$&$\frac{58475}{838656 }$&$\frac{58475}{1490944 }$&$\frac{58475}{4100096 }$&$\frac{58475}{11927552}$ \\ \hline
9         &-       &$\frac{21659}{1677312 }$&$\frac{21659}{1723904 }$&$\frac{1969}{186368 }$&$\frac{21659}{3354624 }$&$\frac{21659}{7268352}$ \\ \hline
10        &-       &$\frac{3}{2560 }$&$\frac{9}{7744 }$&$\frac{9}{8192 }$&$\frac{3}{3328 }$&$\frac{3}{5120}$ \\ \hline
11        &-       &$\frac{1}{22528 }$&$\frac{15}{338944 }$&$\frac{15}{346112 }$&$\frac{5}{124928 }$&$\frac{1}{30720}$ \\ \hline
12        &- &0 &0 &0 &0 &0 \\ \hline
13        &- &0 &0 &0 &0 &0 \\ \hline
14        &- &0 &0 &0 &0 &0 \\ \hline
15        &- &0 &0 &0 &0 &0 \\ \hline 
    \end{tabular}
\end{center} 
\end{table}

\begin{table}
\begin{center}
   \caption{The results for $C_{22}^{~~k}$.
            We show the result from the comparison between the 
two-point function
            $G_{22}$ and the one-point function $F_{1}$ in the second column, 
            and those from the comparison between the three-point functions 
            $H_{22l}$ and the two-point functions $G_{kl}$ in the third column.
            Note that in the latter case, we can use various kinds of 
            observer operators $O_{l}$ labeled by $l$.}
  \label{tab:C22k}
\vspace{5mm}
\def\arraystretch{1.7}
   \begin{tabular}{|c|c|c|c|c|c|c|}  \hline
    &2-point and&\multicolumn{5}{c|}{3-point and 2-point}\\ \cline{3-7}
$k  $ & 1-point & $l=1\sim 6 $&$l= 7 $&$l= 8 $&$l= 9 $&$l= 10$ \\ \hline
1         &$\frac{12}{7    }$&$\frac{12}{7 }$&$\frac{12}{7 }$&$\frac{12}{7 }$&$\frac{12}{7 }$&$\frac{12}{7}$ \\ \hline
2         &-       &$\frac{873}{448 }$&$\frac{873}{448 }$&$\frac{873}{448 }$&$\frac{873}{448 }$&$\frac{873}{448}$ \\ \hline
3         &-       &$\frac{141}{112 }$&$\frac{141}{112 }$&$\frac{141}{112 }$&$\frac{141}{112 }$&$\frac{141}{112}$ \\ \hline
4         &-       &$\frac{15}{32 }$&$\frac{15}{32 }$&$\frac{15}{32 }$&$\frac{15}{32 }$&$\frac{15}{32}$ \\ \hline
5         &-       &- &$\frac{3}{32 }$&$\frac{3}{32 }$&$\frac{3}{32 }$&$\frac{3}{32}$ \\ \hline
6         &-       &- &$\frac{1}{128 }$&$\frac{1}{128 }$&$\frac{1}{128 }$&$\frac{1}{128}$ \\ \hline
7        &-      &$\frac{1440}{637 }$&$\frac{720}{637 }$&$\frac{160}{637 }$&$\frac{1440}{23569 }$&$\frac{1440}{77077}$ \\ \hline
8         &-       &$\frac{298161}{372736 }$&$\frac{33129}{46592 }$&$\frac{298161}{745472 }$&$\frac{298161}{2050048 }$&$\frac{298161}{5963776}$  \\ \hline
9         &-       &$\frac{12665}{69888 }$&$\frac{37995}{215488 }$&$\frac{37995}{256256 }$&$\frac{12665}{139776 }$&$\frac{12665}{302848}$ \\ \hline
10        &-       &$\frac{33}{1280 }$&$\frac{9}{352 }$&$\frac{99}{4096 }$&$\frac{33}{1664 }$&$\frac{33}{2560}$ \\ \hline
11        &-       &$\frac{3}{1408 }$&$\frac{45}{21184 }$&$\frac{45}{21632 }$&$\frac{15}{7808 }$&$\frac{1}{640}$ \\ \hline
12        &-       &$\frac{1}{12288 }$&$\frac{33}{406016 }$&$\frac{33}{409600 }$&$\frac{11}{141312 }$&$\frac{11}{155648}$ \\ \hline
13        &- &0 &0 &0 &0 &0 \\ \hline
14        &- &0 &0 &0 &0 &0 \\ \hline
15        &- &0 &0 &0 &0 &0 \\ \hline 
    \end{tabular}
\end{center} 
\end{table}

On the other hand, 
by comparing the three-point functions and the two-point functions, 
we can determine $C_{nm}^{~~~k}$ for general $k$ through (\ref{eq:threeandtwo}). 
Moreover, by changing the third operator ${\cal O}_l$, 
we can obtain $C_{nm}^{~~~k}$ many times, 
which provides a consistency check on the operator relations 
(\ref{eq:ope}). 
We have calculated $C_{nm}^{~~~k}$ with  $1 \leq n, m \leq 2 $ from 
the relation (\ref{eq:threeandtwo}) for $1 \leq l \leq 10$. 
The results are shown in Tables \ref{tab:C11k}-\ref{tab:C22k}. 
We see that, for $1 \leq  k \leq 6 $, 
the results obtained from different $l$ coincide. 
Note also that the results for $k=1$ coincide with the ones obtained 
by comparing the two-point functions and the one-point function. 

For $k\geq 7$, however, 
the coefficients obtained from different $l$ do not coincide. 
The interpretation of this fact shall be given in the next subsection. 

\subsection{Interpretation of the Results}

In this subsection, we will explain why we obtained 
the consistent results for $k \leq 6$ and not for $k \geq 7$
in the previous subsection.

\begin{figure}
\begin{center}
\leavevmode
\epsfbox{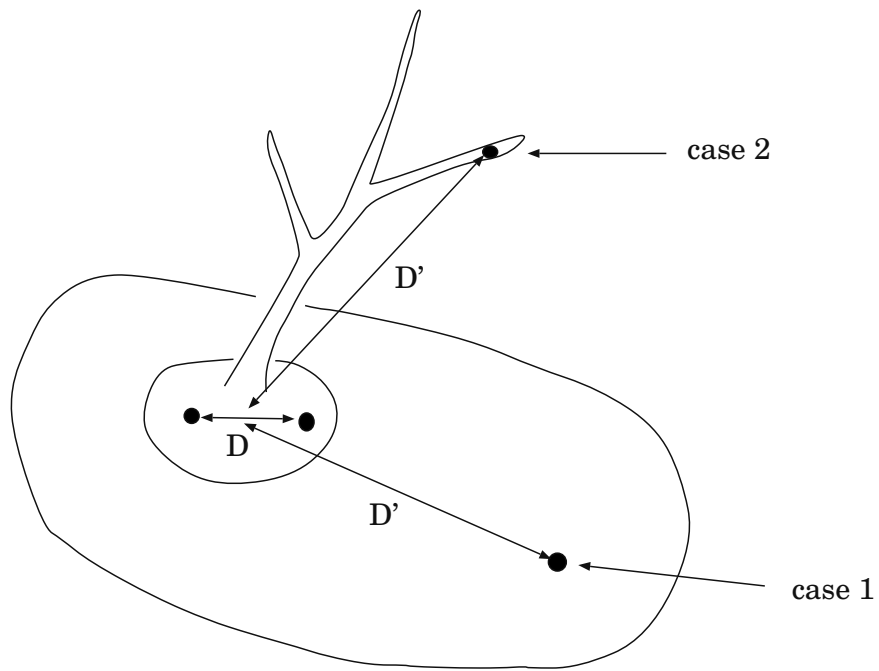}
\caption{The observer operators at the geodesic distance $D'$ 
from the two close operators. 
In case 1 the observer operator can be regarded to be outside the region 
surrounding the two close operators, whereas in case 2 it is at a definite 
distance $D'$ because of  a large fluctuation of the local degrees of freedom 
in the metric.}
\label{fig:kyokuhi}
\end{center}
\vspace{1cm}
\end{figure}

The operator product expansion in ordinary field theories states that the 
two operators which are close to each other can be viewed as a 
superposition of various single operators. 
The existence of such a property itself comes from the fact that 
the dynamical degrees of freedom surrounding the 
two operators smear out the detail of the local structure. 
When we put other operators to see the operator product expansion,
we put them outside the region surrounding the two operators which are close
to each other in order to ensure that the observer operators see 
what comes out after the integration over the surrounding 
degrees of freedom. 
In ordinary field theories, this can be guaranteed by demanding 
that the observer operators are at a definite distance 
from the two operators. 
In quantum gravity, however, we must be careful,
since the metric is fluctuating. 
There are two cases for the positions of the observer operators at a definite
distance from the two operators.(See Fig. \ref{fig:kyokuhi}):
In case 1 the observer operator can be regarded to be outside the region 
surrounding the two close operators, whereas in case 2 it is at a definite 
distance because of  a large fluctuation of the local degrees of freedom in the 
metric which should be integrated out.
We expect OPE holds in case 1.
As we will show in the remainder of this subsection,
the calculations of the coefficients $C_{nm}^{~~~k}$ for $k\leq 6$
are affected only by configurations corresponding to the case 1, 
whereas those for $k\geq 7$ are affected by configurations of both the case 1 
and the case 2.
Therefore we can expect that OPE holds true for $k\leq 6$ 
but not for $k \geq 7$. 
We consider this as a natural explanation for the results 
in the previous subsection.  

\begin{figure}
\begin{center}
\leavevmode
\epsfbox{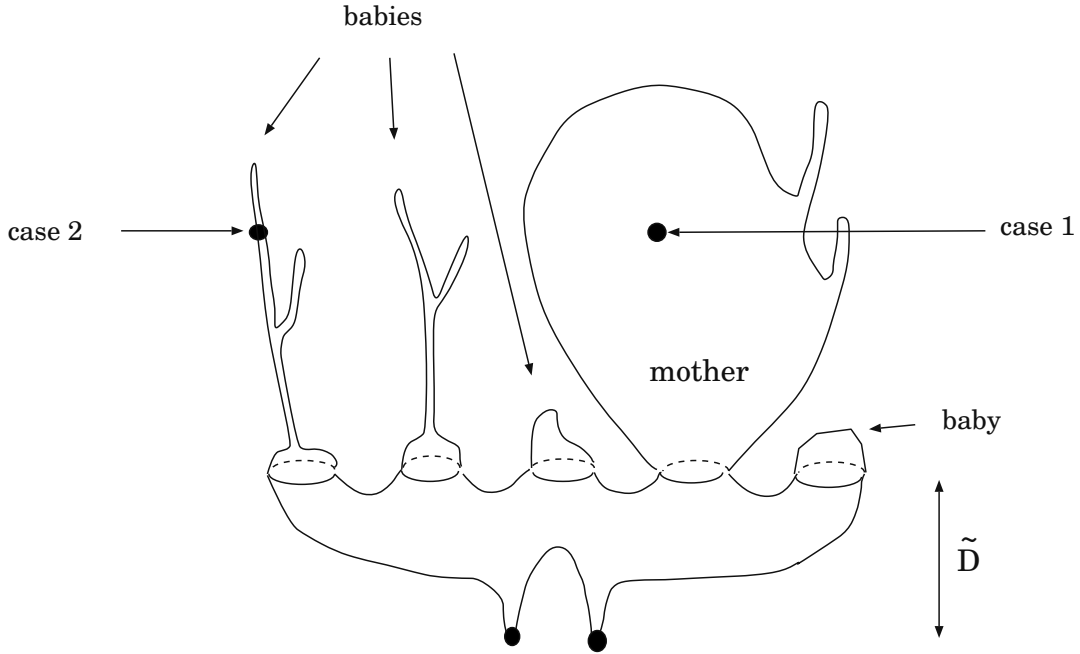}
\caption{Two cases for the positions of the observer operators.
         The section of the surface at the geodesic distance 
         $\tilde{D}$ from the union of the two close operators 
         is composed of many loops, one of which is attached to the mother 
         universe. 
         The case 1 corresponds to the situation where the observer operator 
         lies in the mother universe, 
         while in the case 2 it lies in one of the baby universes.}
\label{fig:baloon}
\end{center}
\vspace{1cm}
\end{figure}

Let us then see how the above claim can be made.
First, we will distinguish the two cases mentioned above definitely as follows.
Let us consider a section of the surface at a geodesic distance 
$\tilde{D}$ from the union of the two close operators 
(See Fig. \ref{fig:baloon}).
We can take an arbitrary value for $\tilde{D}$ if it is greater than $D/2$
and smaller than $D'$.
The section is composed of many loops. 
It is known that only one of the loops is attached to the mother 
universe (infinite-volume universe) and the others are attached to 
baby universes (finite-volume universes) \cite{Kawai}. 
The case 1 corresponds to the situation where the observer operator lies 
in the mother universe, while in the case 2 it lies in one of the baby universes.

\begin{figure}
\begin{center}
\leavevmode
\epsfbox{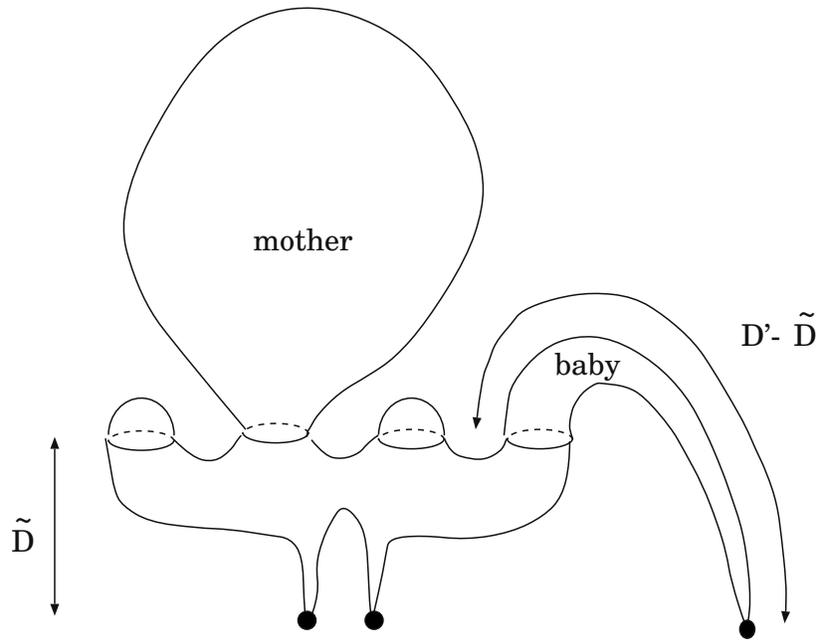}
\caption{Configurations in the three-point functions corresponding to the case 2.
         The observer operator $O_{l}$ lies in a baby universe.}
\label{fig:kobu}
\end{center}
\vspace{1cm}
\end{figure}

As in the previous subsection, let us consider the comparison between 
the three-point functions and the two-point functions.
The case 2 corresponds to the configuration in Fig. \ref{fig:kobu},
where the observer operator ${\cal O}_l$ is in a baby universe.
As we explained in Section \ref{sec:fractal}, taking the thermodynamic limit 
corresponds to taking the term proportional to $t^{3/2}$.
Since the contribution from the mother universe part is proportional to 
$t^{3/2}$, those from baby universe parts should be proportional to 
$t^{0}$.
Hence, from dimensional analysis the baby universe part 
where the observer operator lies in Fig. \ref{fig:kobu} gives the factor 
\beq
G_{jl}(D'-\tilde{D};t=0) \sim (D'-\tilde{D})^{-1-j-l} ~~~~~(j=1,2,3,\cdots).
\eeq
Therefore the configurations corresponding to the case 2 
give only the power of $D'$ lower than $D'{}^{-2-l}$.
As we can see from (\ref{eq:threeandtwo}) and (\ref{eq:GnmTD}),
these configurations only affects the calculations of  
the coefficients $C_{nm}^{~~~k}$ for $k\geq 7$.   

Also in the general case in which 
we obtain $C_{nm}^{~~~k}$ by comparing the $N$-point functions and 
the $(N-1)$-point functions, 
we can show by dimensional analysis 
that the contribution of the configurations with some of the observers 
belonging to baby universes only affects the coefficients 
for $k\geq 7$.\footnote[2]{
Note also that since our arguments are independent of the value of $\tilde{D}$,
the coefficients for $k \leq 6$ are affected only by the configurations
where the observer operators always lie in the mother universe
for arbitrary value of $\tilde{D}$.}

In this way, we have explained why we obtained the consistent OPE coefficients
for $k \leq 6$ and not for $k \geq 7$. 
\vspace{1cm}

\section{Summary and Discussion}
\setcounter{equation}{0}
\hspace*{\parindent}
In this paper, we have calculated the correlation functions with 
fixed geodesic distances up to three-point functions. 
We found that there are scaling operators
${\cal O}_n$ with even $n$, which can not be seen unless the geodesic distance
is fixed.
Elucidating why these new operators appear is an open problem.   
From the two-point functions of the cosmological constant terms with 
fixed geodesic distances, we 
were able to see the fractal structure of the space-time in a more 
direct way than it was seen through the loop-length distribution. 
We examined the OPE in quantum gravity, namely
if two operators close to each other can be 
viewed as a superposition of operators when seen from a distance. 
By comparison of three-point and two-point functions, as well as 
by comparison of two-point and one-point functions, we obtained 
the OPE coefficients $C_{nm}^{~~~k}$, 
which are found to be consistent for $k \le 6$. 

This is because the calculations of the coefficients $C_{nm}^{~~~k}$ for 
$k \leq 6$
are affected only by the configurations where the observer operators lie
in the mother universe.
On the contrary, we obtained inconsistent results for $k \geq 7$
because they are affected by both configurations where the observer operators 
lie in the mother and the baby universes.

There are several things that need to be clarified further.
First of all, we should calculate $N$-point functions with $N\geq4$ in order 
to check further the consistency of the operator product expansion
given in this paper. 
We should see if we can make the $C_{nm}^{~~~k}$ consistent for $k\geq 7$ 
as well by restricting all the observers to be in the mother universe.
Considering higher genus and including matter degrees of freedom are also
interesting extensions of our analysis.  

We expect that operator product expansion holds in the case 
of higher-dimensional quantum gravity as well.
We hope that this kind of approach will eventually enable us 
to understand the universality class of quantum gravity. 

\vspace{1cm}

We would like to thank M. Oshikawa, M. Ikehara and N. Ishibashi 
for stimulating discussion. 
We are also grateful to N.D. Hari Dass for carefully reading the manuscript.

\end{document}